\documentclass[twocolumn,showpacs,preprintnumbers,amsmath,amssymb,prb]{revtex4}
\usepackage{amsmath,amsfonts,bm,graphicx}
\usepackage{color}

\begin{document}

\title{Failure of standard approximations of the exchange coupling in nanostructures}

\author{Jesper Pedersen,$^1$ Christian Flindt,$^1$ Niels Asger Mortensen,$^1$ and Antti-Pekka Jauho$^{1,2}$}
\affiliation{$^1$MIC -- Department of Micro and Nanotechnology,
             NanoDTU, Technical University of Denmark, Building 345east,
             DK-2800 Kongens Lyngby, Denmark\\
             $^2$Laboratory of Physics, Helsinki University of Technology, P.\ O.\ Box 1100, FI-02015 HUT, Finland}

\date{\today}

\begin{abstract}
We calculate the exchange coupling for a double dot system using a
numerically exact technique based on finite-element methods and an
expansion in 2D Gaussians. Specifically, we evaluate the exchange
coupling both for a quasi-one and a two-dimensional system, also
including an applied magnetic field.  Our numerical results provide
a stringent test of standard approximation schemes (e.g.,
Heitler--London, Hund--Mulliken, Hubbard), and they show that the
standard methods do not have reliable predictive power even for simple
model systems. Their value in modeling more realistic quantum-dot
structures is thus cast in serious doubt.
\end{abstract}



\maketitle

\section{Introduction}
The possibility of coherent manipulation of electron spins in
low-dimensional nanostructures, aimed at future large-scale quantum
information processing,\cite{Loss:1998} calls for a thorough
understanding of the spin interactions at play. In the proposal for
quantum computing with quantum dots by Loss and DiVincenzo,
 the exchange coupling between the spins of electrons in
tunnel-coupled quantum dots was envisioned as the controllable
mechanism for coherent manipulation of spin
qubits.\cite{Loss:1998,Burkard:1999} Recently, this fundamental
building block of a possible future solid-state quantum computing
architecture was realized in an experiment, demonstrating
electrostatic control of the exchange coupling.\cite{Petta:2005}

In this paper we present numerically exact finite-element methods
for calculations of the exchange coupling between electron spins in
tunnel-coupled quasi-one and two-dimensional quantum dots. Such
structures have already been under intensive theoretical
investigation using various numerical methods, e.g.\ based on an
exact diagonalization of the underlying Hamiltonian
\cite{Szafran:2004,Bellucci:2004,Helle:2005,Zhang:2006,Melnikov:2006,Melnikov:2006a,Zhang:2007}
or using quantum-chemical approaches like self-consistent
Hartree--Fock methods.\cite{Hu:2000} Such numerical approaches often
require extensive numerical work. Therefore, much attention has been
devoted to simple approximations which lead to closed-form analytic
expressions for the exchange
coupling.\cite{Burkard:1999,Calderon:2006,Saraiva:2007} It is,
however, not immediately obvious to what extent these approximations
yield correct predictions, and where they break down. For example,
in a recent work\cite{Calderon:2006} the validity criterion for such
approximations was the requirement that the exchange coupling at
zero magnetic field always must be positive. A criterion like this
can only provide a necessary condition for an approximate scheme to
be acceptable.

\begin{figure}[b]
\begin{center}
\includegraphics[width=0.4\textwidth, trim = 0 40 0 40,
clip]{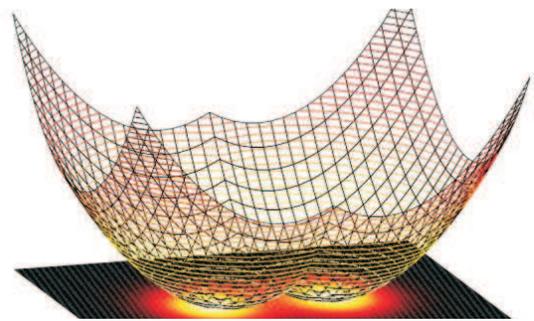}\end{center} \caption{(Color online). Double
quantum dot and numerically calculated charge density. The double
quantum dot is described by the potential $V(\mathbf{r})$ given in
Eq.\ (\ref{eq:potential}) (here with $\alpha=1$,
$\hbar\omega_0=4$~meV and $d/r_0=1$). The two-dimensional contour
plot shows how the charge of two electrons in a singlet spin state
is distributed within the double quantum dot. With finite
tunnel-coupling between the two quantum dots, the spins of the
electrons interact due to the exchange interaction.}
 \label{fig:splash}
\end{figure}

The aim of this work is to provide a quantitative
comparison of the Heitler--London, the Hund--Mulliken, and the
Hubbard approximations, applied to a simple model potential of a
double quantum dot, with numerically exact results.
In particular, we focus on the case, where the distance between the
two quantum dots is short, such that the single-dot electron wave functions
have a large overlap. For short distances, the exchange coupling can
reach values on the order of several meV, making it sufficiently
large to exploit and observe in experiments, and our comparative
study is thus highly relevant for on-going experimental activities
within the field. The finite-element methods used here allow an easy
implementation using available numerical packages,\cite{comsol} also
when finite magnetic fields are included, which strongly influence
the exchange coupling in two-dimensional geometries. We find that
the approximative schemes may provide reasonable predictions of the
exchange coupling for certain parameter ranges, while they fail,
also qualitatively, for short distances, even for the simple model
potential considered here. Their value in modeling more realistic
quantum-dot structures used in experiments is thereby cast in
serious doubt.

\section{Double quantum dot model}

Experimentally, electrons can be confined in double quantum dots
using metallic gates on top of a semiconductor
heterostructure\cite{Petta:2005,Koppens:2005,Hatano:2005} or across
a nanowire\cite{Fasth:2005,Shorubalko:2007} or a
nanotube.\cite{Sapmaz:2006,Jorgensen:2006} By suitable electrostatic
gating such Coulomb-blockade double quantum dots can be brought into
a few-electron regime,\cite{vanderwiel:2003} where only a single
electron occupies each of the two quantum dots. In this regime, the
spin and charge dynamics are described by a two-electron Hamiltonian
of the form
\begin{equation}
H(\mathbf{r}_1,\mathbf{r}_2) = h(\mathbf{r}_1)+
h(\mathbf{r}_2)+C(|\mathbf{r}_1-\mathbf{r}_2|),
 \label{eq_hamiltonian}
\end{equation}
where
\begin{equation}
C(|\mathbf{r}_1-\mathbf{r}_2|)=\frac{e^2}{4\pi\varepsilon_r\varepsilon_0|\mathbf{r}_1-\mathbf{r}_2|}
\end{equation}
is the Coulomb interaction and the single-particle Hamiltonians are
\begin{equation}
h(\mathbf{r})=\frac{\mathbf{p}^2}{2m}+V(\mathbf{r}),
\end{equation}
with $V(\mathbf{r})$ denoting the confining potential. As in many
realizations of double quantum dots we assume that the motion of the
electrons is restricted to maximally two dimensions, i.e.,
$\mathbf{r}=(x,y)$. The inclusion of a magnetic field is discussed
below.

As an illustrative example\footnote{We have applied our methods to
other double dot potentials encountered in the
literature,\cite{Burkard:1999,Szafran:2004} and found that the
reliability of the approximative schemes analyzed in this work are
highly sensitive to the details of the confining potential.} we
consider a simple double dot potential
reading\cite{Helle:2005,Helle:2006}
\begin{equation}
V(\mathbf{r})=\frac{m\omega_0^2}{2}\left[\min\{(x-d)^2,(x+d)^2\}+(\alpha
y)^2\right].
\label{eq:potential}
\end{equation}
Here, $m$ is the effective electron mass, $\hbar\omega_0$ is the
characteristic confinement energy, $2d$ measures the center to
center distance between the quantum dots, and $\alpha$ denotes the
ratio of the confinement strengths in the $x$ and $y$ directions.
Moreover, we introduce the characteristic oscillator length
$r_0=\sqrt{\hbar/m\omega_0}$. The potential is shown in Fig.\
\ref{fig:splash} together with a numerically calculated charge
density. In the limit $d\rightarrow 0$, the potential reduces to
that of a single quantum dot. In our calculations we use material
parameters typical of GaAs ($m=0.067m_e$, $\varepsilon_r=12.9$). We
consider both the quasi-one dimensional limit $\alpha\gg 1$ and the
two-dimensional case $\alpha=1$.

The exchange coupling between the two electrons is a purely orbital
effect which arises as a consequence of the Pauli principle and the
Coulomb interaction which lead to a splitting $J=E_A-E_S$ of the
lowest eigenvalue $E_S$ corresponding to a symmetric orbital
wavefunction of the two electrons,
$\Psi_S(\mathbf{r}_1,\mathbf{r}_2)=\Psi_S(\mathbf{r}_2,\mathbf{r}_1)$,
and the lowest eigenvalue $E_A$ corresponding to an anti-symmetric
orbital wavefunction, $\Psi_A(\mathbf{r}_1,\mathbf{r}_2)=-
\Psi_A(\mathbf{r}_2,\mathbf{r}_1)$. Due to the Pauli principle the
orbital part of a singlet state must be symmetric, while the orbital
part of a triplet state must be anti-symmetric. The splitting of the
orbital wavefunctions may thereby be mapped onto an effective spin
Hamiltonian, $\mathcal{H}=J\,\mathbf{S}_1\!\cdot\!\mathbf{S}_2$. \cite{ashcroft:1976}
The task is to calculate the exchange coupling $J$ as function of
various parameters, e.g., the distance between the quantum dots and
the applied magnetic field. A magnetic field only affects the
exchange coupling significantly in two-dimensional geometries and we
consequently concentrate on the inclusion of a magnetic field in the
two-dimensional case $\alpha=1$.

\section{Validity of approximate methods }
\subsection{Quasi-one dimensional limit}
We first consider the quasi-one dimensional limit $\alpha\gg 1$,
which may be relevant, e.g., for describing confined electrons in
nanowires. In this limit we integrate out the motion in the
$y$-direction and consider an effective one-dimensional model
reading
\begin{equation}
H = h(x_1)+h(x_2)+\widetilde{C}_ {\alpha}(|x_1-x_2|),
\label{eq_1Dproblem}
\end{equation}
where the single-electron Hamiltonian is
\begin{equation}
h(x)=\frac{p_x^2}{2m}+V(x),
\end{equation}
\begin{equation}
V(x)=\frac{m\omega_0^2}{2}[\min\{(x-d)^2,(x+d)^2\}],
\end{equation}
and we have introduced
\begin{equation}
\widetilde{C}_{\alpha}(|x|)=\frac{e^2}{4\pi\varepsilon_r\varepsilon_0}\sqrt{\frac{\alpha}{2\pi
r_0^2}}e^{\alpha x^2/4r_0^2}K_0(\alpha x^2/4r_0^2)
\end{equation}
as the (regularized) Coulomb interaction in one dimension. Here, $K_0$ is
the zeroth-order modified Bessel function of the second kind. The
exchange coupling can now be calculated using finite-elements by
mapping the one-dimensional two-particle problem onto an effective
two-dimensional single-particle problem: We consider the
two-particle wavefunction $\Psi(x_1,x_2)$ as describing a single
fictitious particle with spatial coordinates $\mathbf{\tilde{r}}=(x_1,x_2)$ and
momentum $\mathbf{\tilde{p}}=(p_{x_1},p_{x_2})$. Mathematically, the corresponding
single-particle-like Hamiltonian then reads
\begin{equation}
H=\mathbf{\tilde{p}}^2/2m+W(\mathbf{\tilde{r}}),
\end{equation}
where
\begin{equation}
W(\mathbf{\tilde{r}})=V(x_1)+V(x_2)+\widetilde{C}_{\alpha}(|x_1-x_2|)
\end{equation}
is the effective external potential that the fictitious particle experiences.

In this reformulation of the problem, the symmetry of the original
two-particle wavefunction enters via the boundary condition along
the diagonal $x_2=x_1$. Symmetric wavefunctions fulfill
$\Psi_S(x_1,x_2)=\Psi_S(x_2,x_1)$ and consequently
$\partial_{x_1}\Psi_S(x_1,x_2)|_{x_2=x_1}=\partial_{x_2}\Psi_S(x_1,x_2)|_{x_2=x_1}$
(Neumann condition), while anti-symmetric wavefunctions fulfill
$\Psi_A(x_1,x_2)=-\Psi_A(x_2,x_1)$ and thus
$\Psi_A(x_1,x_2)|_{x_2=x_1}=0$ (Dirichlet
condition).\footnote{Similar boundary conditions have been used in
discussions on the fundamental understanding of identical
particles.\cite{Leinaas:1977}} Since $W(\mathbf{\tilde{r}})$ is a
confining potential, eigenfunctions go to zero in the limit
$|\mathbf{\tilde{r}}|\rightarrow\infty$. In the numerical
calculations we assume that the eigenfunctions are zero outside a
certain  finite range, and we check that the results converge with
respect to an increase of this range. Thus, we only need to solve a
one-particle problem on a finite-size two-dimensional domain with
well-defined boundary conditions. This class of problems is
computationally cheap with available finite-element method
packages.\cite{comsol}

Before discussing the numerical results we briefly review the
standard approximations.\cite{Burkard:1999} In the Heitler--London
approximation the exchange splitting is calculated as
$J_{\mathrm{HL}}=\langle -|H|-\rangle-\langle +|H|+\rangle$ with the
Heitler--London wavefunctions $|\pm\rangle =
(|L\rangle_1|R\rangle_2\pm|R\rangle_1|L\rangle_2)/\sqrt{2(1\pm
|\langle L|R\rangle|^2)}$, where $H$ is the full two-particle
Hamiltonian, and $|L\rangle$ and $|R\rangle$ are the single-particle
Fock--Darwin ground states of a single quantum dot centered at
$\mathbf{r}_L=(-d,0)$ and $\mathbf{r}_R=(d,0)$, respectively. The
Heitler--London approximation can be improved by including doubly
occupied spin singlet states and diagonalizing the Hamiltonian in
the resulting Hilbert space. This is known as the Hund--Mulliken
approach and yields the expression
$J_{\mathrm{HM}}=V-U_r/2+\frac{1}{2}\sqrt{U_r^2+16t_r^2}$. Here,
$U_r$ and $t_r$ are the on-site Coulomb interaction and the tunnel
coupling, respectively, renormalized by the inter-dot Coulomb
interaction as described in Ref.\ \onlinecite{Burkard:1999}, while
$V$ (not to be confused with the confinement potential)
is the difference in Coulomb energy between the singly occupied
singlet and triplet states. Additional details about the
approximative methods are given in Appendix \ref{app:approx}.

If the inter-dot Coulomb interaction is negligible, the renormalized
quantities $U_r$ and $t_r$ reduce to their bare values, $U$ and $t$,
while $V=0$, and if moreover $t/U\ll 1$, the Hund--Mulliken
expression reduces to the standard Hubbard expression
$J_{\mathrm{H}}=4t^2/U$. The Hubbard approximation, which always
predicts a positive exchange energy, obviously cannot explain that
the exchange energy with an applied magnetic field can be negative.
This failure can be corrected by retaining the inter-dot Coulomb
interaction, and in the limit $t_r/U_r\ll 1$, the Hund--Mulliken
approximation then yields the extended Hubbard approximation:
$J^*_{\mathrm{H}}=4 t_r^2/U_r+V$. The energy difference $V$ is
important for the prediction of the exchange coupling at finite
magnetic fields, where it allows for the predicted exchange coupling
to become negative.

\begin{figure}
\begin{center}
\includegraphics[width=0.45\textwidth, trim = 0 0 0 0,
clip]{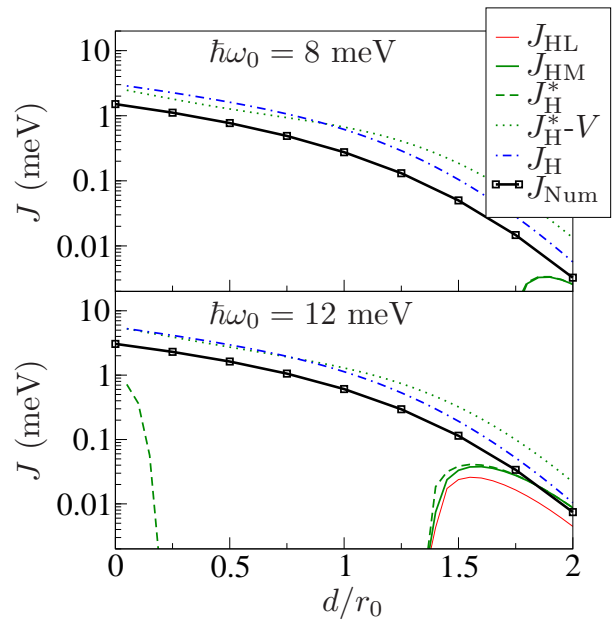}
\end{center}
\caption{(Color online). Exchange coupling as function of
interdot distance in a quasi-one dimensional double quantum dot, $\alpha=10$. The distance $d$
is measured in units of $r_0\equiv\sqrt{\hbar/m\omega_0}$. Together
with the numerical results we show the exchange couplings obtained
with the  Heitler--London $J_{\mathrm{HL}}$, the Hund--Mulliken
$J_{\mathrm{HM}}$, the extended Hubbard $J^*_{\mathrm{H}}$, and the
standard Hubbard $J_{\mathrm{H}}$ approximations. We also show
$J^*_{\mathrm{H}}-V$, where $V$ is the Coulomb energy difference
between the singly occupied singlet and triplet states.} \label{fig:1Dresults}
\end{figure}

In Fig.\ \ref{fig:1Dresults} we show numerical results for the
exchange coupling as a function of the inter-dot distance with
different values of the confinement energy $\hbar\omega_0$ for the quasi-one
dimensional case $\alpha= 10\gg 1$. Together with the numerical results
we show the Heitler--London, the Hund--Mulliken and different
variations of the Hubbard approximations. The validity of the Heitler--London
approximation is strongly dependent on dimensionality due to the increasingly dominating Coulomb
interaction in lower-dimensional systems,\cite{Calderon:2006} and for the quasi-one dimensional case
$J_{\mathrm{HL}}$ is negative in the entire range considered for $\hbar\omega_0 \leq 8$ meV. The
standard Hubbard approximation predicts reasonably well the $d$-dependence, while both the
Hund--Mulliken and extended Hubbard approaches lead to (unphysical) negative values of the exchange coupling for
a wide range of system parameters. We discuss these discrepancies in more detail when we
consider the two-dimensional case below.
Confinement energies larger than $18$ meV are required for these
approximations to yield positive exchange couplings for all interdot distances. For higher values of $\alpha$,
corresponding to stronger confinement in the $y$-direction, the range of
validity of these approximations is further reduced.

\subsection{Two-dimensional case}

We next consider the two-dimensional case $\alpha=1$. In two
dimensions the exchange coupling is strongly dependent on applied
magnetic fields, and we include  a magnetic field perpendicular to
the motion of the electrons by the substitution
$\mathbf{p}\rightarrow\mathbf{p}+e\mathbf{A}$, where
$\mathbf{A}=B_z(-y,x)/2$ is a vector potential corresponding to the
applied magnetic field $B_z \mathbf{\hat{z}}$. The Zeeman term does
not affect the exchange coupling and is trivial to include in final
total energy calculations.

Rather than mapping the two-dimensional two-particle problem onto an
effective four-dimensional one-particle problem, we construct a
two-particle basis from single-particle eigenstates
$\phi_i(\mathbf{r})$ with eigenenergies $\varepsilon_i$ found by
diagonalizing the single-particle Hamiltonian
$h(\mathbf{r})=\frac{(\mathbf{p}+e\mathbf{A})^2}{2m}+V(\mathbf{r})$,
again using finite-element methods.\cite{comsol} The
(unsymmmetrized) two-particle basis functions are then
$\Psi_{i,j}(\mathbf{r}_1,\mathbf{r}_2)=\phi_i(\mathbf{r_1})\phi_j(\mathbf{r_2})$,
in terms of which the matrix elements of the two-particle
Hamiltonian read
\begin{eqnarray}
[\mathbf{H}]_{ij,i'j'}&=&\langle\Psi_{i,j}|H|\Psi_{i',j'}\rangle \nonumber \\
&=&(\varepsilon_i+\varepsilon_j)\delta_{i,i'}\delta_{j,j'}+\langle\Psi_{i,j}|C|\Psi_{i',j'}\rangle.\label{eq:Hijij}
\end{eqnarray}
The Coulomb matrix elements are evaluated by inserting a set of
two-particle states constructed from orthonormalized Gaussian
single-particle wavefunctions. From the low-energy spectrum of
$\mathbf{H}$ we then obtain the exchange coupling $J$. The details
of this procedure are described in Appendix \ref{app:numerics}.

\begin{figure}
\begin{center}
\includegraphics[width=0.45\textwidth, trim = 0 0 0 0,
clip]{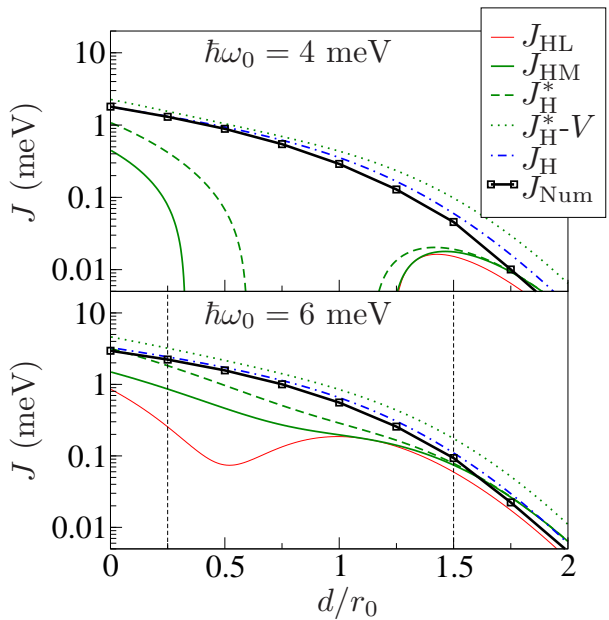}
\end{center}
 \caption{(Color online). Exchange coupling as function of
interdot distance in a two-dimensional double dot, $\alpha=1$.
See Fig.\ \ref{fig:1Dresults} for details. The vertical
lines denote the two values of $d/r_0$ for which the dependence on
the magnetic field is shown in Fig.\ \ref{fig:JvsB2D}.}
\label{fig:Jvsd2D}
\end{figure}

In Fig.\ \ref{fig:Jvsd2D} we show the results for the two-dimensional
case $\alpha=1$. While the standard Hubbard
approximation predicts well the $d$-dependence of the exchange
coupling, the Heitler--London and the Hund--Mulliken approximations
yield predictions that in certain parameter ranges deviate
significantly from the numerical results. In particular, in the case
$\hbar\omega_0 = 4$ meV a range of distances exists around $d=r_0$,
where both approximations predict negative exchange couplings. It is
well-known that the Heitler--London approximation fails at short
distances, when the overlap of the Heitler--London wavefunctions
becomes large, and that the range of validity is reduced as the
ratio between the Coulomb and confinement energy is increased. \cite{Calderon:2006} This explains why the
discrepancies are less pronounced in the case $\hbar\omega_0 = 6$
meV. We conjecture that the poor
predictions by the Hund--Mulliken and the extended Hubbard
approximations are mainly due to the Coulomb energy difference
between the singly occupied singlet and triplet states, denoted $V$,
overestimating the effects of the inter-dot Coulomb interaction at
short distances ($d\sim r_0$), leading to a too low (or even
negative) exchange energy. For large distances ($d\sim 2r_0$), this
overestimation decreases and a better agreement with the full
numerics is obtained. In the figure we also show
$J^*_{\mathrm{H}}-V$ which predicts well the exchange coupling,
indicating that the effects of the inter-dot Coulomb interaction
indeed seem to be overestimated. With larger confinement energies
this overestimation becomes less significant, and a better agreement
with the numerically exact results is found.

\begin{figure}
\begin{center}
\includegraphics[width=0.45\textwidth, trim = 0 0 0 0,
clip]{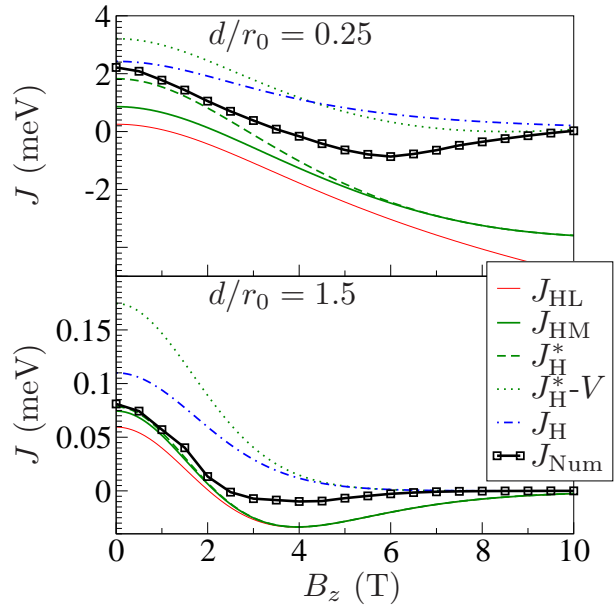}
\end{center}
 \caption{(Color online). Exchange coupling as function of
magnetic field in a two-dimensional double dot.
Results were obtained with $\hbar\omega_0=6$ meV.
See Fig.\ \ref{fig:1Dresults} for details. }
\label{fig:JvsB2D}
\end{figure}

In Fig.\ \ref{fig:JvsB2D} we show numerical results for the exchange
coupling as function of the magnetic field $B$ with different
inter-dot distances $d$. Together with the numerical results we
again show the Heitler--London, the Hund--Mulliken and different
variations of the Hubbard approximations. The results show that {\it
none} of the approximations predict well the dependence of the
exchange coupling over the full range of magnetic fields for short
distances $d<r_0$. For the Hund--Mulliken and the extended Hubbard
approximations we again attribute the discrepancy to an
overestimation of the effects of the inter-dot Coulomb interaction.
For large distances this overestimation is less pronounced, and a
good prediction of the qualitative features is obtained.

\section{Conclusions}

We have presented numerically exact finite-element calculations of
the exchange coupling between electron spins confined in
low-dimensional nanostructures. We have tested a number of
approximations often encountered in the literature by applying them
to a simple double dot potential and found that they only predict
well the exchange coupling in restricted parameter regimes, when
compared to numerical exact results. While the approximative schemes
may yield some insight into the qualitative features of the exchange
coupling, we find it unlikely that they would suffice in the
exchange coupling calculations for actual experimental structures
and experiments, having seen how they may fail even in the case of a
simple model potential.

We thank A.\ Braggio, K.\ Flensberg,  L.\ H.\ Olsen, A.\ S.\ S\o
rensen, and J.\ M.\ Taylor for fruitful discussions. In particular,
we thank A.\ Harju for helpful advice during the development of our
numerical routines.  APJ is grateful to the FiDiPro program of the
Finnish Academy for support during the final stages of this work.

\appendix

\section{Approximative Methods}
\label{app:approx}

In the quasi-one
dimensional limit $\alpha\gg 1$ we have evaluated the approximative
methods numerically using Mathematica, ensuring convergence of the
results with respect to a screening length $\delta\rightarrow 0$ of the
regularized Coulomb interaction. In the following, we list analytical expressions obtained for the various
approximative methods presented in the article for the two-dimensional
case $\alpha=1$.

The single-dot potentials corresponding to the double dot potential
in Eq.\ (\ref{eq:potential}) are those of a harmonic oscillator
centered at $(\pm d,0)$. The single-dot orbitals are thus the
Fock--Darwin states shifted to $(\pm d,0)$. For $d=0$ the Fock--Darwin
ground state is
\begin{equation}
\varphi(x,y)=\sqrt{\frac{m\omega}{\pi\hbar}}e^{-m\omega(x^2+y^2)/2\hbar},
\end{equation}
where $\omega=\omega_0\sqrt{1+\omega_L^2/\omega_0^2}$ with
$\omega_L$ denoting the Larmor frequency $\omega_L=eB/2mc$. In the
presence of a magnetic field given by the vector potential
$\mathbf{A}=B_z(-y,x)/2$, shifting the ground state to $(\pm d,0)$
adds a phase factor of $e^{\pm iyd/2l_B^2}$, where $l_B$ is the
magnetic length $l_B=\sqrt{\hbar c/eB}$. We thus obtain the
single-dot orbitals $\varphi_{\pm d}(x,y) = e^{\pm
iyd/2l_B^2}\varphi(x\mp d, y)$, where $\varphi_{\pm d}(x,y)$ then
denotes the single-dot orbital centered at $(\pm d,0)$.

Using these single-dot orbitals we obtain for the exchange coupling
in the Heitler--London approximation
\begin{eqnarray}
J_\mathrm{HL} &=&
\frac{\hbar\omega_0}{\mathrm{sinh}\left[2d^2(2b-1/b)\right]}
\left[c_s\frac{\sqrt{\pi b}}{2}\left\{e^{-bd^2}I_0(bd^2)\right\}\right.\nonumber \\
&&+\left.\frac{2d}{\sqrt{b\pi}}\left\{1-e^{-bd^2}\right\}+2d^2\left\{1-\mathrm{erf}\left(\sqrt{b}d\right)\right\}\right],\nonumber \\
\end{eqnarray}
where $b$ is the magnetic compression factor $b=\omega/\omega_0$,
$I_0$ is the zeroth-order Bessel function, $\mathrm{erf}(x)$ is the
error function, and we have introduced the dimensionless distance
$d\rightarrow d/r_0$. The prefactor $c$ is the ratio between the
Coulomb and confining energy,
$c_s=\frac{e^2}{4\pi\epsilon_r\epsilon_0r_0}\frac{1}{\hbar\omega_0}$.

In the Hund--Mulliken approximation the exchange coupling is calculated by
diagonalizing the two-electron Hamiltonian in the space spanned by
$\Psi^D_{\pm d}(\mathbf{r}_1,\mathbf{r}_2)=\Phi_{\pm d}(\mathbf{r}_1)\Phi_{\pm d}(\mathbf{r}_2)$
and
$\Psi^S_{\pm}(\mathbf{r}_1,\mathbf{r}_2)=
[\Phi_{+d}(\mathbf{r}_1)\Phi_{-d}(\mathbf{r}_2) \pm  \Phi_{-d}(\mathbf{r}_1)\Phi_{+d}(\mathbf{r}_2)]/\sqrt{2}$,
where $\Phi_{\pm d}$ are the orthonormalized single-particle
states $\Phi_{\pm d} = (\varphi_{\pm d}-g\varphi_{\mp d})/\sqrt{1-2Sg+g^2}$, with
$g=(1-\sqrt{1-S^2})/S$. This leads to the expression
$J_{\mathrm{HM}}=V-U_r/2+\frac{1}{2}\sqrt{U_r^2+16t_r^2}$, where \cite{Burkard:1999}
\begin{eqnarray}
&t_r = t-w = \left< \Phi_{\pm d} \right|h\left| \Phi_{\mp d}\right>
-\left<\Psi_+^S\right|C\left|\Psi_{\pm d}^D\right>/\sqrt{2}, \nonumber \\
&V = V_--V_+ = \left< \Psi_-^S\right|C\left|\Psi_-^S\right> - \left< \Psi_+^S\right|C\left|\Psi_+^S\right>, \nonumber \\
&U_r = U-V_++X = \left<\Psi_{\pm d}^D\right|C\left|\Psi_{\pm d}^D\right> - \left< \Psi_+^S\right|C\left|\Psi_+^S\right> \nonumber \\
& + \left<\Psi_{\pm d}^D\right|C\left|\Psi_{\mp d}^D\right>.
\end{eqnarray}
The Coulomb matrix elements are given by Burkard~\emph{et~al.} in Ref.~\onlinecite{Burkard:1999}
and are applicable to any model potential for which the corresponding
single-dot potential is a simple harmonic oscillator. Thus only the matrix element $t$ is different
for our model potential. We find
\begin{eqnarray}
\frac{t}{\hbar\omega_0} &=& \frac{S}{1-S^2}\left[\frac{d}{\sqrt{\pi b}}\left(1-e^{-bd^2}\right)+d^2\mathrm{erfc}\left(d\sqrt{b}\right)\right],\nonumber \\
\end{eqnarray}
where $\mathrm{erfc}(x)$ is the complimentary error function.

\section{Numerical Methods}
\label{app:numerics}

Here, we discuss the numerical method used in the two-dimensional
case $\alpha=1$. We use finite-element methods to solve the
single-electron problem given by the single-particle Hamiltonian $h$
in Eq.~(\ref{eq_hamiltonian}).\cite{comsol} The full two-electron
problem is then solved by expressing the two-electron Hamiltonian in
Eq.~(\ref{eq_hamiltonian}) in a basis of product states of
single-electron solutions $\left|\psi_n\right>$, in terms of which
the matrix elements are given by Eq.~(\ref{eq:Hijij}).
To evaluate the Coulomb
elements the single-electron eigenstates are expanded in an
orthonormalized basis of 2D Gaussians
$\phi_{n_x,n_y}(x,y)=x^{n_x}y^{n_y}e^{-r^2/2}$, where $n_x$ and
$n_y$ are positive integers or zero. The Coulomb matrix elements
between product states of 2D Gaussians can be determined
analytically, and we state the result here for convenience
\cite{Pedersen:2007}
\begin{widetext}
\begin{eqnarray*}
C_{ijkl} &=& \frac{e^2}{4\pi\epsilon_0\epsilon_r}\;\frac{\pi}{2}
\left(-\frac{1}{4}\right)^{n/2}
\sum_{s_1=0}^{\lfloor n_1/2\rfloor }\cdots\sum_{s_4=0}^{\lfloor n_4/2\rfloor }\left(-1\right)^{n_3+n_4+s_1+s_2-s_3-s_4} \nonumber \\
&&\times\;
\frac{\Gamma\left(n_1+1\right)}{\Gamma\left(s_1+1\right)\Gamma\left(n_1-2s_1+1\right)}\cdots
\frac{\Gamma\left(n_4+1\right)}{\Gamma\left(s_4+1\right)\Gamma\left(n_4-2s_4+1\right)} \nonumber \\
&&\times\;
\frac{\Gamma\left[\left(n_1+n_3-2s_1-2s_3+1\right){ /2}\right]\Gamma\left[\left(n_2+n_4-2s_2-2s_4+1\right){/2}\right]}
{\Gamma\left[\left(n-2s\right){ /2}+1\right]} \times\;
2^{(n-2s+1)/2}\;\Gamma\left[\left(n-2s+1\right)/2\right],
\end{eqnarray*}
\end{widetext}
for $n_1+n_3$ and $n_2+n_4$ even and zero otherwise. Here
$C_{ijkl}=\left<\phi_{n_{x,i},n_{y,i}}\phi_{n_{x,j},n_{y,j}}\right|C\left|\phi_{n_{x,k},n_{y,k}}\phi_{n_{x,l},n_{y,l}}\right>$
while $\Gamma(x)$ is the Gamma function and $\lfloor n/2\rfloor$
indicates flooring of half-integers. Above, we have introduced
$n_1=n_{x,i}+n_{x,k}$, $n_2=n_{y,i}+n_{y,k}$, $n_3=n_{x,j}+n_{x,l}$
and $n_4=n_{y,j}+n_{y,l}$ , while $n=\sum_in_i$ and $s=\sum_is_i$.
The two-particle Hamiltonian matrix resulting from this procedure
may then be diagonalized in the subspaces spanned by the symmetric
and antisymmetric product states, respectively, to yield the
exchange coupling. Because the expansion in 2D Gaussians becomes
increasingly inaccurate as the interdot distance is increased, we
are limited to interdot distances of the order of the characteristic
oscillator length $r_0$. The accuracy of the 2D Gaussian expansion
at larger interdot distances could be greatly improved by using an
expansion in relative coordinates.\cite{Helle:2005}

The finite-element calculations of the single-particle states can be
carried out with very high efficiency using existing finite-element
packages\cite{comsol} and are not a limiting factor in terms of
computational time or convergence. Also, the Coulomb matrix elements
$C_{ijkl}$ may be pre-calculated and saved in a lookup table, such
that the largest portion of the computational time is spent
assembling the two-electron Hamiltonian matrix. For each matrix
element, a total of $N^2$ lookups in the $C_{ijkl}$ table are
required, where $N$ is the number of 2D Gaussians included in the
expansion set. A significant reduction in computational time is
accomplished by utilizing the symmetry of the Hamiltonian in the
product state basis, limiting the calculation to matrix elements
which differ by more than a simple complex conjugation. For the
results presented in this article, a total of 100 2D Gaussians were
used to ensure that the results obtained may essentially be
considered exact. With this basis set and a total of $7^2=49$
single-particle product states, the calculation of the exchange
coupling takes approximately two and a half hours on a standard
computer equipped with an Intel Core2 Duo 1.86MHz CPU. As few as 25
Gaussians are in many cases sufficient to produce results that are
within 10\% of the exact results, and in that case a single
calculation only takes about 5 minutes.

The use of finite-element methods for solving the single-electron
problem makes it easy to construct the two-electron Hamiltonian,
even if analytic expressions for the matrix elements of the
single-electron Hamiltonian in the basis of 2D Gaussians cannot
easily be obtained. This makes the method very flexible, and only
little work is required to solve problems with different choices of
potentials. We have verified our numerical implementation against
the results in Ref.\ \onlinecite{Helle:2005} as well as for the
simple problem of two opposite spin particles in a two-dimensional
parabolic potential, which can be solved analytically.


\end{document}